\documentclass[aps,prd,amssymb,nofootinbib,twocolumn,epsf,floatfix,showpacs]
{revtex4}
\usepackage[usenames]{color} 
\usepackage{ulem} 
\usepackage{graphicx}
\usepackage{amssymb}
\usepackage{amsmath}
\usepackage{epstopdf}

\begin{document}


\title{Solving the Einstein Constraints Numerically on Compact
  Three-Manifolds \\ Using Hyperbolic Relaxation}

\author{Fan Zhang\,${}^{a,b}$ and Lee Lindblom\,${}^c$}

\affiliation{${}^a$Gravitational Wave and Cosmology Laboratory,
    Department of Astronomy, Beijing Normal University, Beijing 100875, China}

\affiliation{${}^b$Advanced Institute of Natural Sciences,
  Beijing Normal University at Zhuhai 519087, China}

  \affiliation{${}^c$Department of Physics, University 
of California at San Diego, La Jolla, CA 92093, USA}

\date{\today}
 
\begin{abstract}
The effectiveness of the hyperbolic relaxation method for solving the
Einstein constraint equations numerically is studied here on a variety
of compact orientable three-manifolds.  Convergent numerical solutions
are found using this method on manifolds admitting negative Ricci
scalar curvature metrics, i.e. those from the $H^3$ and the $H^2\times
S^1$ geometrization classes. The method fails to produce solutions,
however, on all the manifolds examined here admitting non-negative
Ricci scalar curvatures, i.e. those from the $S^3$, $S^2\times S^1$,
and the $E^3$ classes.  This study also finds that the accuracy of the
convergent solutions produced by hyperbolic relaxation can be
increased significantly by performing fairly low-cost standard
elliptic solves using the hyperbolic relaxation solutions as initial
guesses.
\end{abstract}


\maketitle


\section{Introduction}
\label{s:introduction}

Hyperbolic relaxation was introduced by R\"uter, et al.~\cite{Ruter2018}
as a method of solving elliptic partial differential equations
numerically using a hyperbolic evolution code, without the need to
develop a stand alone elliptic solver.  The basic idea is to transform
the elliptic equations into hyperbolic ones whose late time solutions
approach the solutions to the original elliptic problem.  As a simple
example consider the elliptic equation
\begin{equation}
  \tilde\nabla^a\tilde\nabla_a \psi = f(\psi,x^a),
  \label{e:BasicEllipticEq}
\end{equation}
where $\tilde\nabla_a\tilde\nabla^a$ is a covariant Laplace operator,
and $f(\psi,x^a)$ is a function that may depend on the scalar field
$\psi$ and the spatial coordinates $x^a$.  The hyperbolic relaxation
method introduces a non-physical time coordinate, $t$, and transforms
Eq.~(\ref{e:BasicEllipticEq}) into the damped wave equation:
\begin{equation}
  -\partial_t^{\,2}\psi - \kappa\, \partial_t \psi
  + \tilde\nabla^a\tilde\nabla_a\psi = f(\psi,x^a),
  \label{e:BasicDampedWaveEq}
\end{equation}
where $\kappa$ is a damping parameter.  It is easy to show that all
the solutions to the homogeneous, $f(\psi,x^a)=0$, version of
Eq.~(\ref{e:BasicDampedWaveEq}) with $\kappa>0$ (on a domain without
boundary or using outgoing boundary conditions on a domain with
boundary) drive $\partial_t\psi$ toward zero, see
Appendix~\ref{s:Appendix}. Thus $\psi$ approaches a solution to the
original homogeneous elliptic Eq.~(\ref{e:BasicEllipticEq}).  More
generally the hyperbolic relaxation method can be applied to the
inhomogeneous Eq.~(\ref{e:BasicEllipticEq}) in cases where the
solutions to Eq.~(\ref{e:BasicDampedWaveEq}) drive $\partial_t\psi$
toward zero.  This method has been used successfully to solve the
Einstein constraint equations numerically for black hole
spacetimes~\cite{Ruter2018, Assumpcao2022}.

This paper explores the use of the hyperbolic relaxation method for
solving the Einstein constraint equations on compact orientable
three-manifolds.  A few solutions to the Einstein constraints were
found numerically on a variety of these compact manifolds using
standard elliptic numerical methods in Ref.~\cite{Zhang2022}.  Those
elliptic methods (e.g. using the ksp linear solver and the snes
non-linear solver from the PETSC software
library~\cite{petsc-user-ref}) were found to be very inefficient.
This inefficiency severely limited the ability to find solutions on
compact orientable manifolds having topologies that required
complicated multicube structures to represent them~\cite{Lindblom2013,
  Lindblom2022}.  Hyperbolic relaxation transports constraint
violations throughout the computational domain more efficiently than
the diffusive processes used by many elliptic solvers. Our motivation
for this study was to determine whether hyperbolic relaxation could be
useful for overcoming those inefficiency problems when solving the
Einstein constraint equations for initial data on compact orientable
initial surfaces.

The particular form of the Einstein constraint equation used in this
study has a simple form belonging to the class of elliptic equations
in Eq.~(\ref{e:BasicEllipticEq}):
\begin{eqnarray}
  \tilde\nabla^a\tilde\nabla_a\psi &=& \textstyle\frac{1}{8}\psi\,
  \left(\tilde R - \psi^4 \langle \tilde R\,\rangle\right),
  \label{e:CMCConstraintSAlt}
\end{eqnarray}
where $\tilde\nabla_a$ is the covariant derivative and $\tilde R$ the
scalar curvature determined by a positive definite metric $\tilde
g_{ab}$. The constant $\langle\tilde R\,\rangle$ is the spatial
average of the scalar curvature,
\begin{equation}
  \langle \tilde R\, \rangle
  = \frac{\int \sqrt{\det\tilde g}\,\tilde R\, d^{\,3}x}
  {\int \sqrt{\det\tilde g}\, d^{\,3}x}.
  \label{e:AvarageRtildeDef}
\end{equation}
The physical meaning of $\psi$ determined by
Eq.~(\ref{e:CMCConstraintSAlt}) is the conformal factor needed to
transform $\tilde g_{ab}$ into the Einstein initial value constraint
satisfying physical spatial metric $g_{ab}$,
\begin{eqnarray}
  g_{ab} &=& \psi^4 \tilde g_{ab},
  \label{e:ConformalMetric}
\end{eqnarray}
in a vacuum spacetime with cosmological constant $\Lambda$ on
a spacelike surface with mean curvature $K$ given by
\begin{equation}
  K^2 = 3\Lambda - \textstyle\frac{3}{2}\langle \tilde R\, \rangle,
  \label{e:CMCParameterChoices}
\end{equation}
see Ref.\cite{Zhang2022} for details.  The conformal factor, $\psi$,
determined by Eq.~(\ref{e:CMCConstraintSAlt}) also has the interesting
mathematical property that it transforms $\tilde g_{ab}$ into a metric
$g_{ab}$ whose scalar curvature $R$ is constant:
\begin{equation}
  R = \langle \tilde R\,\rangle.
\label{e:ConstantScalarR}
\end{equation}
Thus $\psi$ is a solution to the Yamabe problem~\cite{Yamabe1960} that
constructs a constant scalar curvature geometry on the manifold.

The numerical methods used in this study to solve the hyperbolic
relaxation Eq.~(\ref{e:BasicDampedWaveEq}) are described in
Sec.~\ref{s:SolvingHyperbolicEquations}.  Numerical results of using
these methods on a variety of compact orientable three-manifolds are
described in Sec.~\ref{s:NumericalResults}.  An elliptic refinement
method for improving the accuracy of the numerical solutions found by
hyperbolic relaxation is described in
Sec.~\ref{s:EllipticRefinements}.  This elliptic refinement method
performs a standard elliptic solve with fairly lax convergence
criteria using the hyperbolic relaxation results as its initial guess.
The successes and failures of the tests reported here are summarized
and discussed in Sec.~\ref{s:Discussion}.

\section{Solving Hyperbolic Relaxation Equations Numerically}
\label{s:SolvingHyperbolicEquations}

Numerical solutions to the hyperbolic relaxation version of the
Einstein constraint Eq.~(\ref{e:CMCConstraintSAlt}) are studied here
on a collection of compact orientable three-manifolds.  These
manifolds are represented as multicube structures~\cite{Lindblom2013},
consisting of a collection of cubic regions $\mathcal{B}_A$ whose
faces $\partial_\alpha\mathcal{B}_A$ are identified with its
neighbors' faces $\partial_\beta\mathcal{B}_B$ by a collection of maps
$\Psi^{A\alpha}_{B\beta}$.  The particular way the cubic regions are
glued together by these maps determines the topologies of the
manifolds represented in this way.

The differentiable structures of multicube manifolds are determined by
a reference metric $\tilde g_{ij}$ that (together with the maps
$\Psi^{A\alpha}_{B\beta}$) determines the continuity of vector and
tensor fields across the interface boundaries between regions.  In
particular these reference metrics can be used to construct the
outward directed unit normal vectors $\tilde n^i$ at each point on
each face of each cubic region.  The differentiable structure ensures
the outward directed unit normals are identified with the
corresponding inward directed normals on the adjoining faces of
neighboring regions.  The $C^1$ reference metrics for the manifolds
included in this study were constructed as described in
Ref.~\cite{Lindblom2015}. The collection of Cartesian coordinate
charts in the cubic regions of the multicube structure, including its
boundary identification maps $\Psi^{A\alpha}_{B\beta}$ and the
differential structure provided by the reference metric $\tilde
g_{ij}$, serves as a global atlas of coordinate charts for these
manifolds.  These reference metrics are also used as the conformal
metrics in the solutions to the Einstein constraint equations studied
here.

The hyperbolic relaxation version of the Einstein constraint
Eq.~(\ref{e:CMCConstraintSAlt}) is given by
\begin{eqnarray}
  -\partial_t^{\,2}\psi - \kappa\, \partial_t \psi
  + \tilde\nabla^a\tilde\nabla_a\psi &=& \textstyle\frac{1}{8}\psi\,
  \left(\tilde R - \psi^4 \langle \tilde R\,\rangle\right).
  \label{e:CMCConstraintHRForm}
\end{eqnarray}
This equation is solved numerically by converting it into the
first-order symmetric hyperbolic system,\footnote{R\"uter, et
    al.~\cite{Ruter2018} use a slightly different first-order
    representation of the Einstein constraints, see Eqs.~(17)--(19) of
    their paper.}
\begin{eqnarray}
  \partial_t\psi &=& -\Pi,
  \label{e:psiDot} \\
  \partial_t\Pi + \tilde g^{ij}\tilde\nabla_i\Phi_j &=& -\kappa\,\Pi+
  \textstyle\frac{1}{8}\psi\,
  \left(\tilde R - \psi^4 \langle \tilde R\,\rangle\right),
  \label{e:PiDot}  \\
  \partial_t\Phi_i + \tilde\nabla_i\Pi  &=&
  \gamma_2 \left(\tilde\nabla_i\psi-\Phi_i\right).
  \label{e:PhiDot}
\end{eqnarray}
The auxiliary first-order field $\Pi$ is equivalent to
$-\partial_t\psi$ from Eq.~(\ref{e:psiDot}), while $\Phi_i$ is
equivalent to $\tilde\nabla_i\psi$ when the constraint
$\mathcal{C}_i\equiv\Phi_i-\tilde\nabla_i\psi=0$ is satisfied.  The
constant $\gamma_2>0$ ensures that Eq.~(\ref{e:PhiDot}) damps away any
constraint violations that may be introduced during the evolution.
This first-order symmetric hyperbolic representation of the hyperbolic
relaxation equation has the same principal parts, and therefore the
same characteristic fields, as the system introduced in
Ref.~\cite{Holst2004}.  The propagating characteristic fields $U^\pm$
with characteristic speeds $\pm 1$ are given by
\begin{equation}
  U^{\pm}=\Pi\pm \tilde n^i\Phi_i - \gamma_2 \psi,
\end{equation}
where $\tilde n^i$ is the outward directed unit normal vector at the
boundary.  As in any first-order symmetric hyperbolic system, boundary
conditions are imposed on the incoming characteristic fields ($U^-$ in
this case) at each boundary point of each cubic region.  At the
interface boundaries between cubic regions these incoming fields are
set to the values of the outgoing fields ($U^+$ in this case) copied
from the corresponding boundary points of the neighboring region.

\section{Numerical Results}
\label{s:NumericalResults}

Numerical solutions to the hyperbolic relaxation version of the
Einstein constraint Eq.~(\ref{e:CMCConstraintHRForm}) have been
computed in this study for a collection of ten compact orientable
three-manifolds.  These manifolds, listed in
Table~\ref{t:3-manifold_list}, represent examples from five of the
eight Thurston geometrization classes.  The manifold names used here
are those from Ref.~\cite{Lindblom2015}, which includes the multicube
structures of each manifold and explains in detail how the reference
metrics, $\tilde g_{ij}$, are constructed.
Table~\ref{t:3-manifold_list} also lists the spatial average of the
Ricci curvature $\langle\tilde R\,\rangle$, defined in
Eq.~(\ref{e:AvarageRtildeDef}), the spatial volume
$\tilde{\mathcal{V}}$, defined by
\begin{equation}
  \tilde{\mathcal{V}} = \int \sqrt{\det \tilde g}\, d^{\,3}x,
  \label{e:tildeVDef}
\end{equation}
and the Thurston geometrization class for each manifold.  

\begin{table}[!h]  
    \scriptsize \renewcommand{\arraystretch}{1.4}
\begin{center}
  \caption{Compact orientable three-manifolds used in this study to
    test the hyperbolic relaxation method for solving the Einstein
    constraint equations.
    \label{t:3-manifold_list}}
  \begin{ruledtabular}
  \begin{tabular}{l c c c}
    Manifold & $\langle\tilde R\, \rangle$ & $\tilde{\mathcal{V}}$
    & Geometrization    \vspace{-0.14cm} \\
    &&& Class     \vspace{0.07cm} \\
    \hline
    $G2\times S1$ & -2.9676 & 15.287 & $H^2\times S^1$ \\
    $G5\times S1$ & -2.9676 & 61.149 &$H^2\times S^1$ \\
    Seifert-Weber Space & -4.7990 & 27.327 & $H^3$ \\
    Sixth-Turn Space (E5) & $0.0028$ & 11.126 & $E^3$ \\
    $KB/n2 \times\!\!\sim S1$ & 1.2715 & 48.946 & $E^3$ \\
    $SFS[RP2.n2:(2,1)(2,-1)]$ & 1.2715 & 48.946 & $E^3$ \\
    $S2\times S1$ & 2.6878 & 16.768 & $S^2\times S^1$ \\
    $S3$ & 5.8899 & 36.592 & $S^3$ \\
    $L(10,3)$ & 1.7134 & 28.029 & $S^3$  \\
    $SFS[S2:(2,1)(2,1)(2,-1)]$ & 2.6552 & 19.150 & $S^3$ \\
  \end{tabular}
  \end{ruledtabular}
\end{center}
\end{table}

The hyperbolic relaxation Eqs.~(\ref{e:psiDot})--(\ref{e:PhiDot}) were
solved using the multicube coordinate systems described in
Ref.~\cite{Lindblom2022} for these manifolds.  These representations
consist of Cartesian coordinate charts within a collection of
non-overlapping cubic regions that intersect only at the interfaces
between neighboring cubes.  The hyperbolic relaxation equations were
solved using pseudo-spectral methods on Gauss-Lobatto collocation
points in each cubic region.  The time evolutions were performed using
an eighth-order Dormand-Prince integrator with the error tolerance set
to $10^{-12}$.  Boundary conditions were imposed at the interfaces
between cubic regions using the multi-penalty method.

The Einstein constraint damping parameter, $\kappa$, that appears in
Eq.~(\ref{e:CMCConstraintHRForm}), was set in these tests to the
value,
\begin{equation}
  \kappa = \frac{2\pi}{{\tilde{\mathcal{V}}^{1/3}}},
  \label{e:kappaDef}
\end{equation}
where $\tilde{\mathcal{V}}^{1/3}$ represents a characteristic length
scale of the manifold.  The first-order scalar-wave constraint damping
parameter, $\gamma_2$, was set to the relatively large value
$\gamma_2=100$ in these tests to ensure that violations of the
scalar-wave constraint, $\mathcal{C}=\Phi_i-\tilde\nabla_i\psi$, were
strongly suppressed.

\begin{figure}[!h]
\centerline{\includegraphics[width=3in]{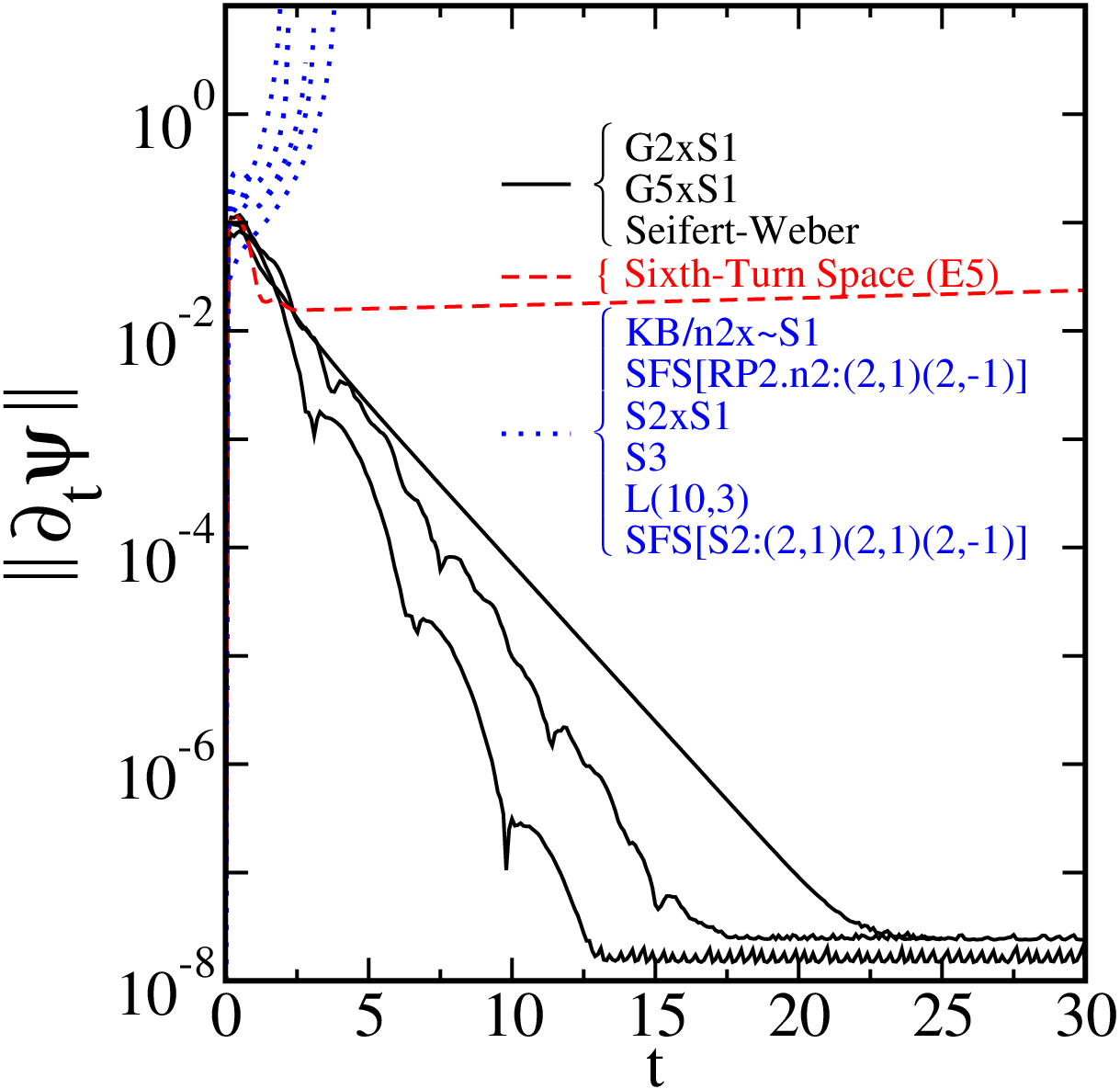}}
\caption{\label{f:PsiDot}.  The norm $||\partial_t\psi||$ evolves
  toward zero for the $G2\times S1$, $G5\times S1$, and the
  Seifert-Weber manifolds, indicating that hyperbolic relaxation
  successfully solves the Einstein constraints asymptotically on these
  manifolds.  However, $||\partial_t\psi||$ grows without bound on the
  other manifolds studied here, so hyperbolic relaxation fails to find
  solutions to the Einstein constraints on those manifolds.}
\end{figure}

Figure~\ref{f:PsiDot} shows the time dependence of
$||\partial_t\psi||$, the $L_2$ norm of $\partial_t\psi$, defined by
\begin{equation}
  ||\partial_t\psi||^2 = \frac{1}{\mathcal{\tilde V}}
    \int \left(\partial_t\psi\right)^2
  \sqrt{\det \tilde g}\, d^{\,3}x,
  \label{e:PsiDotNormDef}
\end{equation}
for the numerical evolutions of
Eqs.~(\ref{e:psiDot})--(\ref{e:PhiDot}) on each of the manifolds
listed in Table~\ref{t:3-manifold_list}.  The solid (black) curves
represent the evolutions on the $G2\times S1$, $G5\times S1$ and the
Seifert-Weber manifolds.  Those evolutions reduce $||\partial_t\psi||$
to small values at late times, so hyperbolic relaxation is successful
in computing approximate solutions to the Einstein constraint equation
at late times for those cases. The evolutions shown as dashed (red) or
dotted (blue) curves represent the evolutions on the remaining
manifolds in this study.  Those evolutions show $||\partial_t \psi||$
growing exponentially (or faster), thus hyperbolic relaxation fails to
produce solutions to the Einstein constraint equation in those cases.
Appendix~\ref{s:Appendix} provides an analysis of the stability of
these hyperbolic relaxation evolutions. That analysis shows that
hyperbolic relaxation evolutions are likely to be stable on compact
manifolds with $\langle\tilde R\,\rangle\leq0$, and unstable on those
manifolds with $\langle\tilde R\,\rangle>0$, which is consistent with
these numerical results.

\begin{figure}[!h]
\centerline{\includegraphics[width=3in]{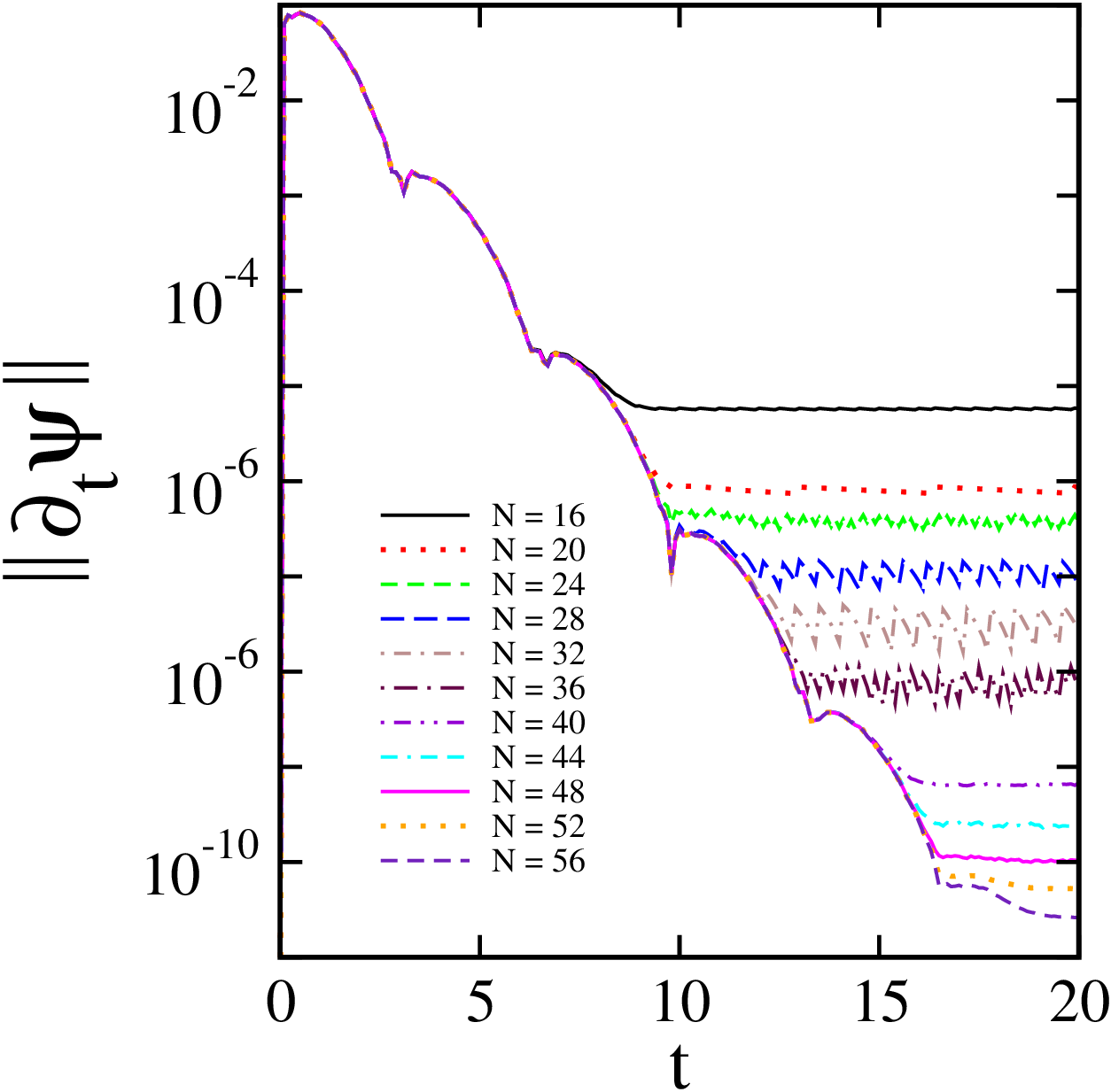}}
\caption{\label{f:PsiDotConvergence}.  The evolutions of the norm
  $||\partial_t\psi||$ at different spatial resolutions, $16\leq N
  \leq 56$, for hyperbolic relaxation evolutions on the Seifert-Weber
  manifold.  The late time convergence of $||\partial_t\psi||$ toward
  zero as $N$ is increased demonstrates that hyperbolic relaxation
  does produce numerically convergent solutions to the Einstein
  constraint equation for this case.}
\end{figure}

The evolutions shown in Fig.~\ref{f:PsiDot} were computed at the same
numerical resolution for each manifold: $N=35$ grid points in each
dimension in each cube of the multicube structure.
Figure~\ref{f:PsiDotConvergence} illustrates $||\partial_t\psi||$ for
the evolutions on the Seifert-Weber manifold using a sequence of
numerical resolutions in the range $16\leq N \leq 56$.  These results
demonstrate that $||\partial_t\psi||$ converges toward zero at late
times as the resolution $N$ is increased.  The results for evolutions
on the $G2\times S1$ and the $G5\times S1$ manifolds have similar late
time convergences, so those graphs are not included here.

\begin{figure}[!h]
\centerline{\includegraphics[width=3in]{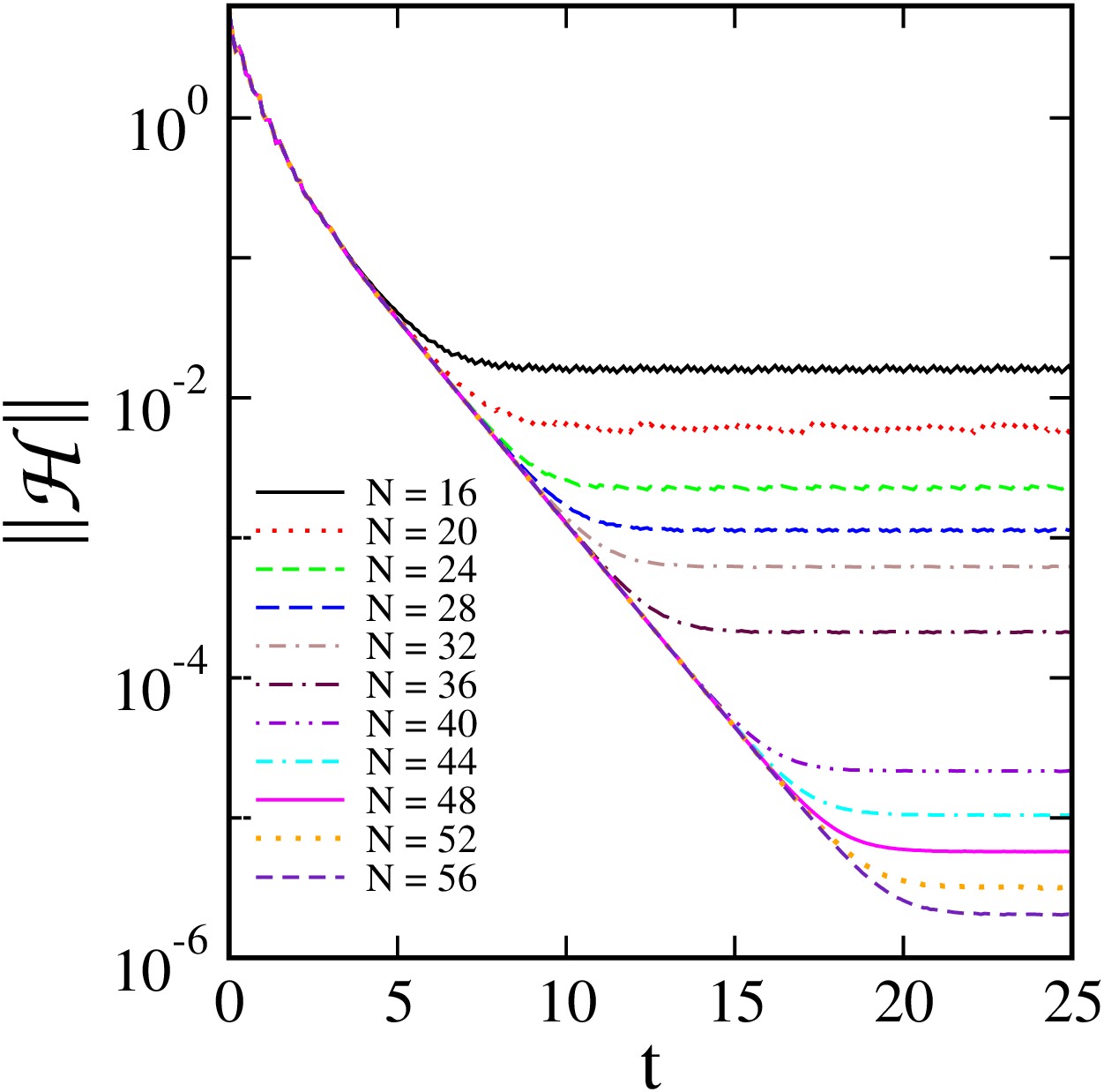}}
\caption{\label{f:G2xS1Evolutions}.  The Hamiltonian constraint norm
  $||\mathcal{H}||$ as a function of time, $t$, for hyperbolic
  relaxation evolutions on the $G2\times S1$ manifold.}
\end{figure}

\begin{figure}[!h]
\centerline{\includegraphics[width=3in]{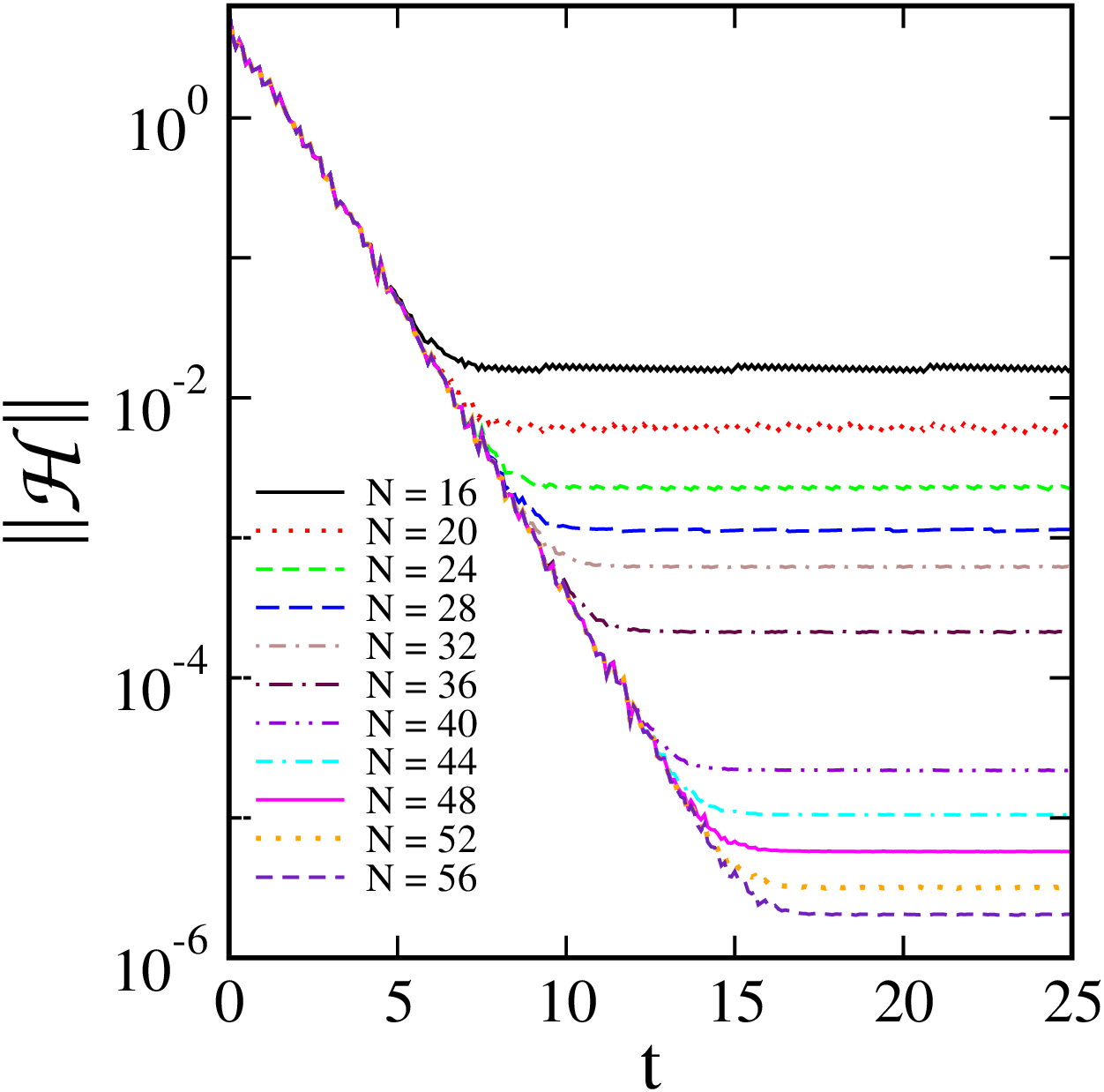}}
\caption{\label{f:G5xS1Evolutions}.  The Hamiltonian constraint norm
  $||\mathcal{H}||$ as a function of time, $t$, for hyperbolic
  relaxation evolutions on the $G5\times S1$ manifold.}
\end{figure}

\begin{figure}[!h]
\centerline{\includegraphics[width=3in]{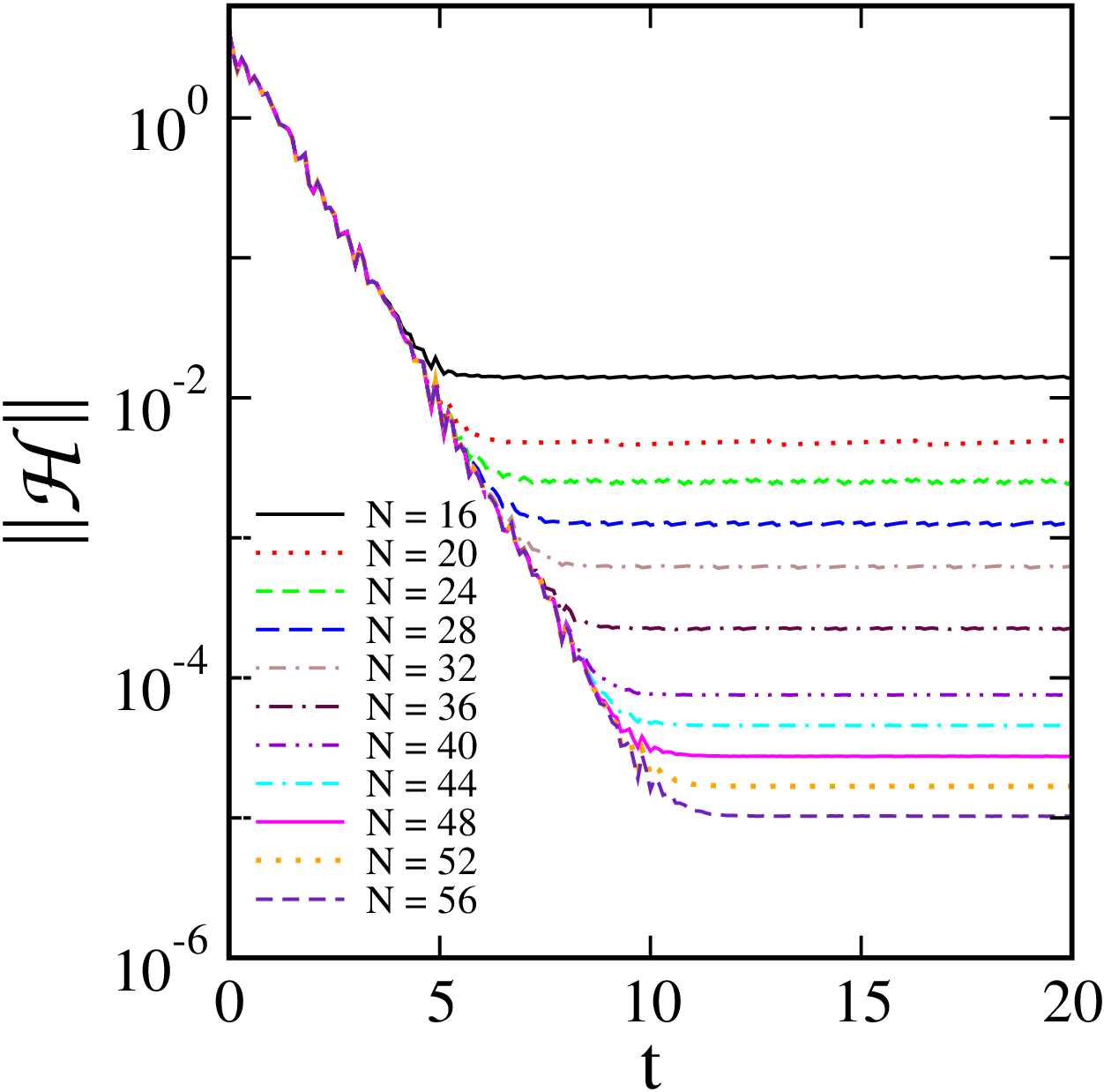}}
\caption{\label{f:SeifertWeberEvolutions}.  The Hamiltonian constraint norm
  $||\mathcal{H}||$ as a function of time, $t$, for hyperbolic
  relaxation evolutions on the Seifert-Weber manifold.}
\end{figure}

The Hamiltonian constraint $\mathcal{H}$ for the initial value problem
described in Eq.~(\ref{e:CMCConstraintSAlt}) can be written in its
original geometric form:
\begin{equation}
  \mathcal{H}=R - \langle \tilde R\,\rangle,
  \label{e:HamiltonianConstraint}
\end{equation} 
where $R$ is the Ricci scalar curvature computed from the physical
metric $g_{ij}$ defined in Eq.~(\ref{e:ConformalMetric}).  The constraint
norm $||\mathcal{H}||$ defined by
\begin{equation}
  ||\mathcal{H}||^2 = \frac{1}{\mathcal{\tilde V}}
  \int \mathcal{H}^2
  \sqrt{\det \tilde g}\, d^{\,3}x,
  \label{e:HamiltonianConstraintNorm}
\end{equation}
is a quantitative measure of how well the original Einstein
constraints are satisfied.  Figures~\ref{f:G2xS1Evolutions},
\ref{f:G5xS1Evolutions}, and \ref{f:SeifertWeberEvolutions} illustrate
the time dependence of $||\mathcal{H}||$ for the hyperbolic relaxation
evolutions on the $G2\times S1$, $G5\times S1$, and Seifert-Weber
manifolds respectively using a range of numerical resolutions $16\leq
N \leq 56$.

\begin{figure}[!h]
\centerline{\includegraphics[width=3in]{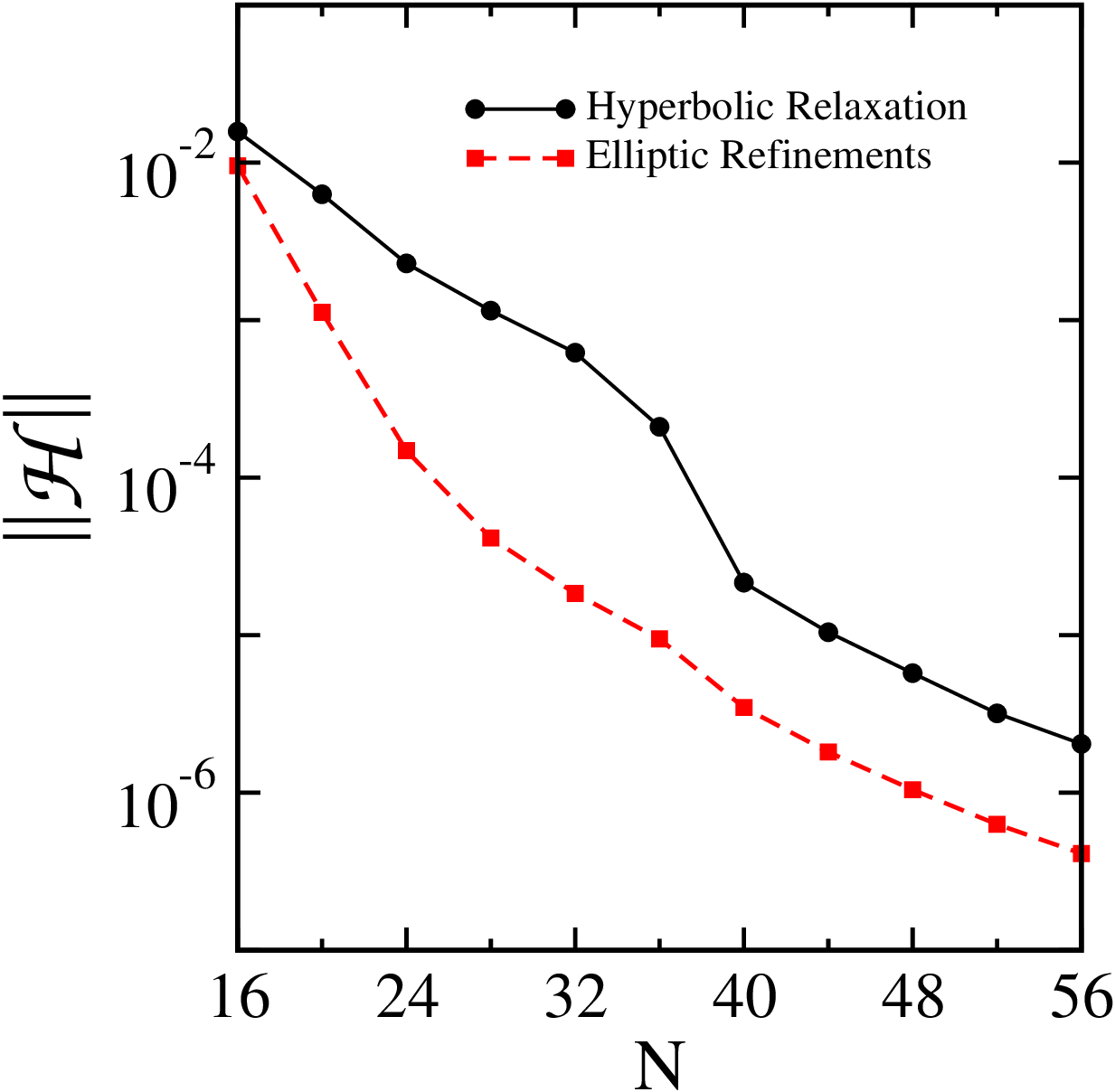}}
\caption{\label{f:G2xS1Convergence} The Hamiltonian constraint norm
  $||\mathcal{H}||$ as a function of the spatial resolution $N$ on the
  $G2\times S1$ manifold.  The results from hyperbolic relaxation
  evolutions (evaluated at the time $t=25$) are represented as the
  solid (black) curve and the elliptic refinements of these results
  are represented as the dashed (red) curve.}
\end{figure}

\begin{figure}[!h]
\centerline{\includegraphics[width=3in]{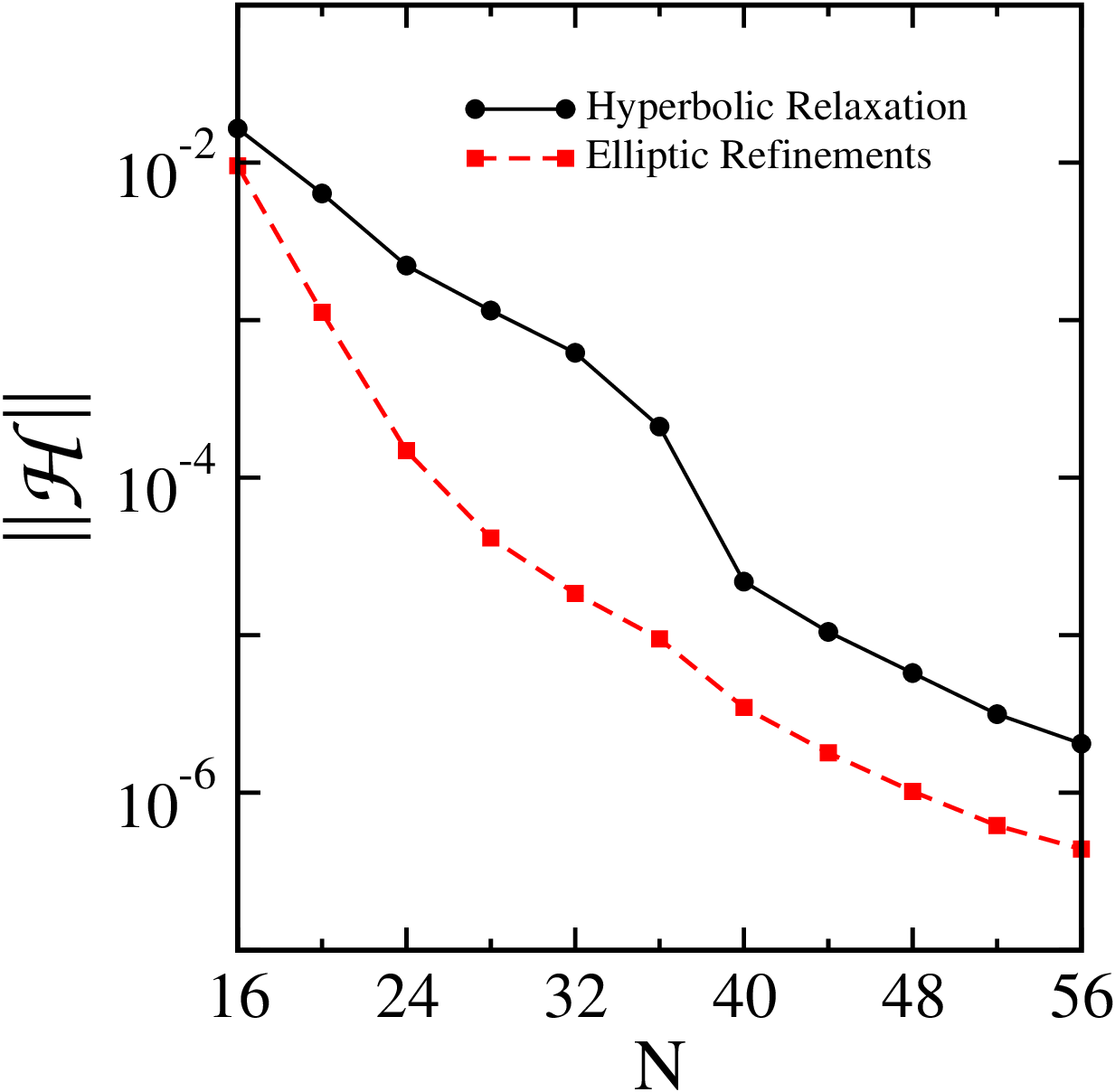}}
\caption{\label{f:G5xS1Convergence} The Hamiltonian constraint norm
  $||\mathcal{H}||$ as a function of the spatial resolution $N$ on the
  $G5\times S1$ manifold.  The results from hyperbolic relaxation
  evolutions (evaluated at the time $t=25$) are represented as the
  solid (black) curve and the elliptic refinements of these results
  are represented as the dashed (red) curve.}
\end{figure}

\begin{figure}[!h]
\centerline{\includegraphics[width=3in]{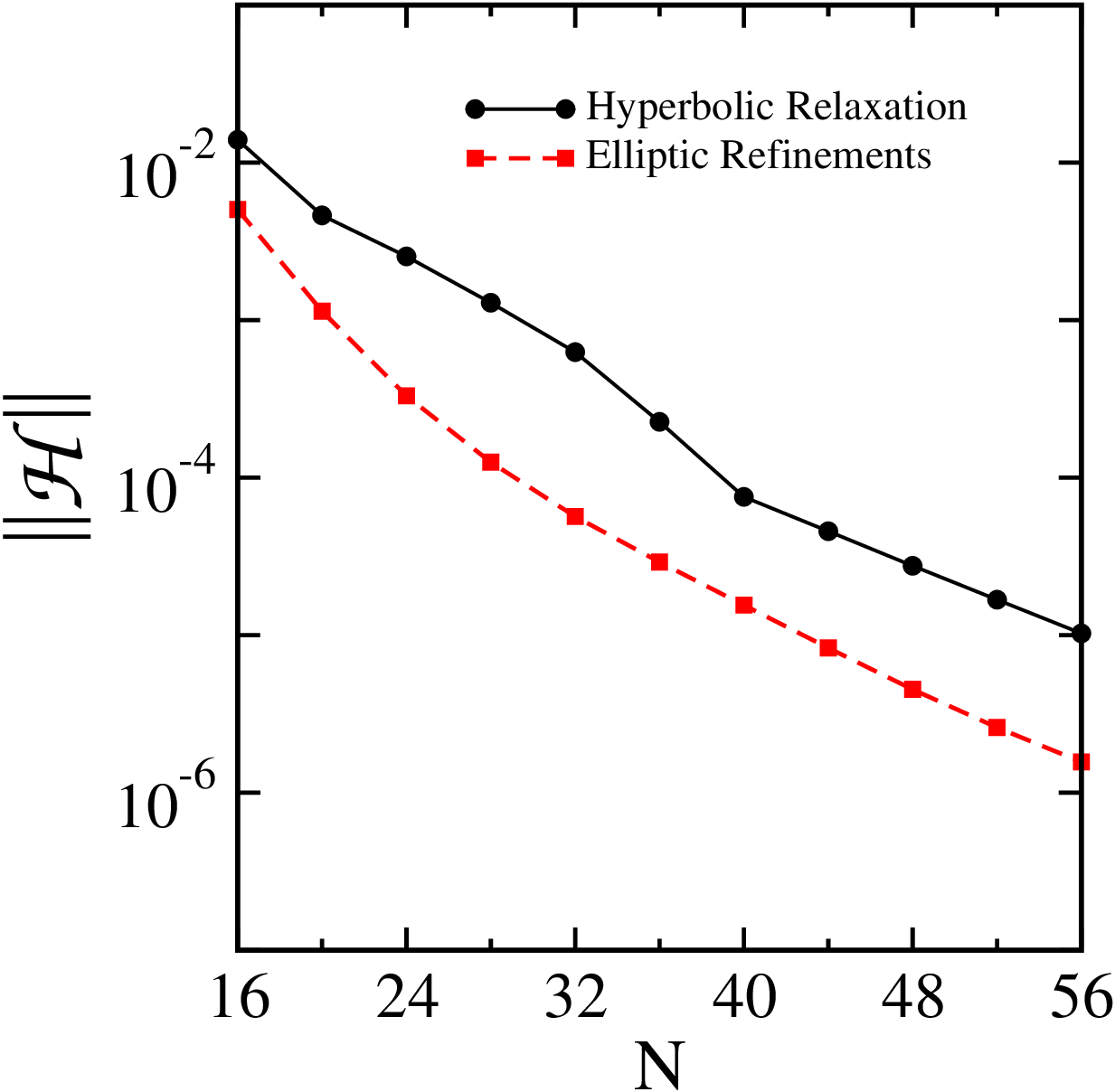}}
\caption{\label{f:SeifertWeberConvergence} The Hamiltonian constraint
  norm $||\mathcal{H}||$ as a function of the spatial resolution $N$
  on the Seifert-Weber manifold.  The results from hyperbolic
  relaxation evolutions (evaluated at the time $t=20$) are represented
  as the solid (black) curve and the elliptic refinements of these
  results are represented as the dashed (red) curve.}
\end{figure}

Each of the evolutions in Figs.~\ref{f:G2xS1Evolutions},
\ref{f:G5xS1Evolutions}, and \ref{f:SeifertWeberEvolutions} show the
Hamiltonian constraint norm $||\mathcal{H}||$ initially decreasing
(approximately) exponentially, and then becoming (essentially) time
independent at late times.  The late time asymptotic values of
$||\mathcal{H}||$ have been extracted from these evolutions at the
time $t=25$ for the evolutions on the $G2\times S1$ and $G5\times S1$
manifolds, and at the time $t=20$ for the Seifert-Weber manifold.
These late time Hamiltonian constraint norms have been plotted as the
solid (black) curves in Figs.~\ref{f:G2xS1Convergence},
\ref{f:G5xS1Convergence}, and \ref{f:SeifertWeberConvergence} as
functions of the spatial resolution $N$.  These results show that
hyperbolic relaxation does produce numerically convergent solutions to
the Einstein constraint equation on the $G2\times S1$, $G5\times S1$
and the Seifert-Weber manifolds.

\section{Elliptic Refinements}
\label{s:EllipticRefinements}

The hyperbolic relaxation evolutions of the Einstein constraint
equation illustrated in
Figs.~\ref{f:G2xS1Evolutions}--\ref{f:SeifertWeberEvolutions} show
that the Hamiltonian constraint norms $||\mathcal{H}||$ approach
resolution dependent constant values at late times.  What determines
these late time values of $||\mathcal{H}||$ in these evolutions?  Are
these simply determined by the truncation errors in the numerical
representations of the conformal factor $\psi$?  Or, is some other
numerical effect setting a higher error floor for the values of the
late time constraint norms?  In an effort to understand these
questions we performed standard numerical elliptic solves (using the
ksp linear solver and the snes non-linear solver from the PETSC
software library) using the final hyperbolic relaxation results as
initial guesses.  The error tolerance parameters for these elliptic
solves were set at fairly large values to keep the cpu run times
reasonably short.  The resulting values of the Hamiltonian constraint
norms $||\mathcal{H}||$ for these elliptic refinement computations are
shown as the dashed (red) curves in
Figs.~\ref{f:G2xS1Convergence}--\ref{f:SeifertWeberConvergence}.  The
resulting $||\mathcal{H}||$ from the elliptic refinement computations
are about an order of magnitude smaller than those from the hyperbolic
relaxation evolutions.  This means that numerical truncation error is
not the limiting factor in the accuracy of the hyperbolic relaxation
evolutions.

While we do not fully understand why elliptic refinement is able to
improve the hyperbolic relaxation results so significantly.  However,
the graphs of $||\partial_t\psi||$ in Fig.~\ref{f:PsiDotConvergence}
might provide some insight.  Our numerical implementation of the
hyperbolic relaxation equations does not drive $||\partial_t\psi||$ to
zero, or even to double precision roundoff levels.  This means that
the hyperbolic relaxation evolutions of the conformal factor $\psi$
never get to completely time independent states.  The oscillations in
$||\partial_t\psi||$ seen in some of the resolutions in
Fig.~\ref{f:PsiDotConvergence} suggest that the second time
derivatives $\partial_t^{\,2}\psi$ may be even larger than
$\partial_t\psi$ in some cases.  The fact that $\partial_t \psi$ and
$\partial_t^{\,2}\psi$ do not vanish in the late time numerical
hyperbolic evolutions provides one mechanism that could prevent the
Hamiltonian constraint norm $||\mathcal{H}||$ from reaching truncation
error levels.

We attempted to find a way to suppress the anomalous time dependence
at late times in our numerical solutions to the hyperbolic relaxation
equation (e.g. by adjusting the damping coefficient $\kappa$, reducing
the timestep error tolerance, increasing the scalar-wave constraint
damping coefficient $\gamma_2$, etc.).  But we were not successful in
improving the results reported here.  These results suggest that the
accuracy of the hyperbolic relaxation method could be improved
significantly if a way could be found to drive $\partial_t\psi$ and
$\partial_t^{\,2}\psi$ more effectively to smaller levels at late
times.  Alternatively, using elliptic refinement can be used to
improve the accuracy of those solutions significantly at relatively
low additional computational cost.

\section{Discussion} 
\label{s:Discussion}

This study explores the use of hyperbolic relaxation to solve the
Einstein constraint equations numerically on compact orientable
three-manifolds.  A primary result of this study is that hyperbolic
relaxation evolutions of the Einstein constraints are unstable on
manifolds with positive conformal scalar curvature averages,
$\langle\tilde R\,\rangle > 0$, while those with negative curvatures,
$\langle\tilde R\,\rangle < 0$, are stable.  This result severely
limits the class of manifolds on which hyperbolic relaxation can be
used successfully.

A second important result of this study is that our implementation of
the hyperbolic relaxation method does not produce solutions whose
accuracy is limited by truncation error.  The late time errors in
$\partial_t\psi$ and $\mathcal{H}$ are dominated by their values along
the edges and at a few collocation points near the edges of the cubic
coordinate patches.  These maximum errors converge to zero with
increasing numerical resolution in much the same way as the $L_2$
norms shown in the figures.  Our attempts to reduce these errors
further by adjusting the filtering and the various damping parameters
were not successful.  The fact that these errors do not converge all
the way to truncation error levels shows, however, that our code's
implementation of the hyperbolic boundary conditions along those edges
is not optimal.

Finally, a third important result of this study is the fact that the
accuracy of the hyperbolic relaxation solutions can be improved by
about an order of magnitude by doing a fairly low cost standard
elliptic solve using the hyperbolic relaxation results as initial
guesses.  The elliptic solver imposes a different set of boundary
conditions in a different way than the hyperbolic evolution system.
These results show that the elliptic solver does a better job than the
hyperbolic evolution code of implementing the correct physical
boundary conditions.

\appendix

\section{}
\label{s:Appendix}

This appendix analyzes the stability of the hyperbolic relaxation
method for finding solutions to the Einstein constraint equations.
Let $\psi_0$ denote a solution to the Einstein constraints given in
Eq.~(\ref{e:CMCConstraintSAlt}),
\begin{eqnarray}
  \tilde\nabla^a\tilde\nabla_a\psi_0 &=& \textstyle\frac{1}{8}\psi_0\,
  \left(\tilde R - \psi_0^4 \langle \tilde R\,\rangle\right).
  \label{e:CMCConstraintSAlt2}
\end{eqnarray}
We examine the stability of hyperbolic relaxation evolutions by
defining $\delta\psi=\psi-\psi_0$ and studying the solutions to the
linearized hyperbolic relaxation Eq.~(\ref{e:CMCConstraintHRForm}):
\begin{eqnarray}
  -\partial_t^{\,2}\delta\psi - \kappa\, \partial_t \delta\psi
  + \tilde\nabla^a\tilde\nabla_a\delta\psi &=&
  \textstyle\frac{1}{8}\delta\psi\,
  \left(\tilde R - 5\psi_0^4\, \langle \tilde R\,\rangle\right).\nonumber\\
  \label{e:CMCConstraintHRForm2}
\end{eqnarray}
Multiplying this equation by $\partial_t\delta\psi$ and integrating
over a compact manifold (such as those included in this study)
results in the equation
\begin{equation}
  \frac{dE}{dt} = - 2\int \kappa \left(\partial_t\delta\psi\right)^2
  \sqrt{\det \tilde g}\, d^{\,3}x,
  \label{e:PerturbationEnergyCons}
\end{equation}
where $E$ is a perturbation energy defined by
\begin{eqnarray}
  E &=& \int \left[\left(\partial_t\delta\psi\right)^2
    + \tilde\nabla^a\delta\psi\, \tilde\nabla_a\delta\psi\right.\nonumber\\
    &&\left.\qquad+\textstyle\tfrac{1}{8}\left(\delta\psi\right)^2
    \left(\tilde R - 5\,\psi_0^{\,4}\langle\tilde R\,\rangle\right)\right]
  \sqrt{\det \tilde g}\, d^{\,3}x.\qquad
  \label{e:PerturbationEnergy}
\end{eqnarray}
Equation~(\ref{e:PerturbationEnergyCons}) shows that all
near-equilibrium evolutions of the hyperbolic relaxation equation
drive the perturbation energy $E$ to smaller values.  The kinetic
terms on the first line of the left side of
Eq.~(\ref{e:PerturbationEnergy}) are positive definite, while the
potential-like terms on the second line do not have definite sign.

If the potential-like terms in Eq.~(\ref{e:PerturbationEnergy}) are
non-negative, then $E$ would be positive definite, including for
example the homogeneous case where the right side of
Eq.~(\ref{e:CMCConstraintHRForm2}) vanishes. In this case hyperbolic
relaxation produces stable evolutions that drive $E$ towards its
minimum value.  However, if the potential-like terms are negative,
then $E$ would not be bounded below, and unstable evolutions would
occur.  We note that $\tilde R$ and $\langle\tilde R\,\rangle$ are
related on compact orientable manifolds by the identities
\begin{eqnarray}
  \int\tilde R\sqrt{\det\tilde g}\,d^{\,3}x &=&
  \int\langle\tilde R\,\rangle \sqrt{\det\tilde g}\,d^{\,3}x,
  \label{e:RtildeAvgDef2}\\
  \int\psi_0\tilde R\sqrt{\det\tilde g}\,d^{\,3}x &=&
  \int\psi_0^{\,5}\langle\tilde R\,\rangle \sqrt{\det\tilde g}\,d^{\,3}x.
  \label{e:CurvatureIdentity}
\end{eqnarray}
The first identity is just the definition of $\langle\tilde R\,\rangle$
from Eq.~(\ref{e:AvarageRtildeDef}), while the second is obtained by
integrating Eq.~(\ref{e:CMCConstraintSAlt2}) over the manifold.  These
identities suggest that the following approximate equality should also
be true,
\begin{eqnarray}
  \int\tilde R\sqrt{\det\tilde g}\,d^{\,3}x &\approx&
  \int\psi_0^{\,4}\langle\tilde R\,\rangle \sqrt{\det\tilde g}\,d^{\,3}x.
  \label{e:ApproxCurvatureIdentity}
  \\\nonumber
\end{eqnarray}
For manifolds with reference metrics having positive average scalar
curvatures, $\langle\tilde R\,\rangle>0$, the term proportional to
$\langle\tilde R\,\rangle$ in $E$ is negative definite. Compared to
Eq.~(\ref{e:ApproxCurvatureIdentity}), the additional factor of 5
multiplying the $\langle\tilde R\,\rangle$ term in $E$, should make
that term dominate over the term containing $\tilde R$ in most (if not
all) cases.  In these cases $E$ will not be bounded below and the
hyperbolic relaxation evolutions will be unstable.  Conversely for
manifolds where $\langle \tilde R\,\rangle<0$, the term proportional
to $\langle\tilde R\,\rangle$ in $E$ is positive definite, and
hyperbolic relaxation is likely to be stable.  This analysis of the
stability of the hyperbolic relaxation method is only qualitative.
However, the results of the numerical tests reported in
Sec.~\ref{s:NumericalResults} show that in practice the sign of
$\langle\tilde R\,\rangle$ is a useful predictor of the stability or
instability of this method, as suggested by this analysis.


\acknowledgments

We thank Nicholas Taylor for a number of helpful discussions about the
numerical methods used in this study, and David Hilditch and Hannes
R\"uter for their comments on an earlier draft of this paper.
L.L. thanks the Department of Physics, National Central University,
Zhongli, Taiwan for their hospitality during a visit in which a
portion of this research as completed. F.Z. was supported in part by
the National Natural Science Foundation of China grants 12073005 and
12021003. L.L. was supported in part by the National Science
Foundation grant 2012857 to the University of California at San Diego,
USA.


\bibstyle{prd} 
\bibliography{paper.bbl}
\end{document}